\documentstyle[12pt,aaspp4]{article} 
\begin{document}
\doublespace
\setlength{\oddsidemargin}{0cm}
\setlength{\evensidemargin}{0cm}
\setlength{\textwidth}{15cm}
\setlength{\topmargin}{-1cm}
\setlength{\textheight}{22cm}

\begin{center}
\hfill  ICRR-Report-378-97-1\\
\hfill  RESCEU 3/97\\
\hfill  UTAP-249\\
\hfill  astro-ph/9612237\\

\title{Big Bang Nucleosynthesis
       and Lepton Number Asymmetry\\ in the Universe}

\author{K. Kohri, \ M. Kawasaki}
\affil{Institute for Cosmic Ray Research, University of Tokyo,
Tanashi,  Tokyo 188, Japan}
\authoraddr{kohri@icrr.u-tokyo.ac.jp \\ 
            kawasaki@icrr.u-tokyo.ac.jp}
\and
\author{Katsuhiko Sato}
\affil{Department of Physics and Research Center for the Early
Universe, School of Science, University of
  Tokyo, Tokyo 113, Japan}
\authoraddr{sato@phys.s.u-tokyo.ac.jp}

\end{center}

\begin{abstract}
Recently it has been  reported that there may be a discrepancy between 
big bang nucleosynthesis theory and observations (BBN crisis).
We show that BBN predictions
agree with the primordial
 abundances of  light elements, ${}^4$He,
D, ${}^3$He and ${}^7$Li inferred from the observational
data if the electron neutrino has
 a net chemical potential $\xi_{\nu_e}$ due to lepton asymmetry.
 We estimate that 
$\xi_{\nu_e} = 0.043^{+0.040}_{-0.040} $(95$\%$ C.L.)  and 
$\Omega_bh^2 = 0.015\;^{+0.006}_{-0.003}$
(95$\%$ C.L.).

\end{abstract}
\newpage

\section{Introduction}
Standard big bang cosmology gives us a simple  and natural picture of
our universe since it  explains the origin of the cosmic
microwave background and the abundances of the light elements, ${}^4$He,
D, ${}^3$He and ${}^7$Li. The
standard big bang nucleosynthesis (SBBN) has a 
single free parameter, i.e. the baryon to photon number 
ratio $ \eta \equiv n_{\rm B}/n_{\gamma}$ where $n_{\rm B}$ and
$n_{\gamma}$ are the
number densities of baryons and photons.
 Comparing SBBN predictions with the abundances of light elements 
inferred from observational data, we obtain the baryon density of the
universe, $\Omega_b \approx 0.01$ (\cite{walker}, \cite{kernan}).

However, recently Hata et al. (1995) pointed out that there is a 
 discrepancy between theory and observation  by using detailed
statistical analysis for both observational
data and theoretical prediction. This discrepancy is called BBN
crisis.
Of course if we adopt the large uncertainties of the D data observed
in Ly-$\alpha$ clouds (\cite{songaila}, \cite{ahogan}, \cite{bhogan}), 
BBN predictions agree with the observed abundances
(\cite{olive}, \cite{copi}).
However if we adopt the chemical evolution model of D and ${}^3$He or 
low D data observed in Ly-$\alpha$ clouds (\cite{tytlera}, 
\cite{tytlerb}), the crisis still exists.
In this paper we discuss the problem by taking the discrepancy
 between the theory and observations seriously.
   One way to solve the discrepancy might be to adopt
some modifications of standard physics used in SBBN, for example, a
large chemical potential for a neutrino or an 
 $O(10{\rm MeV})$ mass (\cite{kawasaki}). 
 It is also pointed out that 
an exotic massive decaying particle may destroy D and ${}^3$He
 after the BBN epoch (\cite{holt}). 

Since there is baryon
asymmetry in the universe, it is quite natural to consider the
possibility that there is also lepton asymmetry.\footnote{In the
standard model during electroweak phase transition the large lepton
asymmetry is  converted into baryon asymmetry (\cite{shapo}).
 Then the lepton number density $n_{\rm L}$ is roughly the same as
 baryon number density $n_B \sim 10^{-10}n_{\gamma}$.
 Thus the total lepton number asymmetry 
 is small. However lepton number asymmetry for each species of
 neutrino  may be large.}  
If there exists a lepton asymmetry for
 neutrinos, they will have finite chemical potentials.
One can reduce the abundance of ${}^4$He if
 an electron neutrino has a large chemical potential. 
This neutrino degeneracy was first studied by (\cite{wagoner}). 
So far  many people
 have imposed the constraints on the neutrino chemical
potentials for $\nu_e$ (\cite{sato}, ~\cite{sato84}, ~\cite{sato85},
 ~\cite{reeves},  ~\cite{goret}, ~\cite{sherrer} ~\cite{kim}),
 for $\nu_e$ and $\nu_{\mu}$ (\cite{yahil}, ~\cite{beaudet})
 and for three neutrinos 
$\nu_e$, $\nu_{\mu}$, $\nu_{\tau}$  (\cite{kang}
,  \cite{thomas}). However a precise statistical analysis was not made 
 in the previous work and the observational data have been updated
 since  that time. Therefore in this paper, 
 we study BBN with the effects of the neutrino
degeneracy using  Monte Carlo simulation (\cite{kawano}) and make a 
likelihood analysis (\cite{crisis}) using the most recent 
data. 

\section{BBN with Neutrino Degeneracy and Likelihood Analysis}
 
Theoretical predictions should be compared with the constraints of
observational data. Hereafter we use the notation 
 Y$_{p}$ ( ${}^4$He mass
fraction), $y_{2p}$ = (D/H)$_{p}$, $y_{3p}$ =
 (${}^3$He/H)$_p$, $y_{7p}$ = (${}^7$Li/H)$_p$
, (number fraction relative to H respectively) where ``$p$'' denotes
the primordial. 
For ${}^4$He we adopt the recent analysis of observations
 in low metallicity extragalactic HII
regions, from which primordial abundance 
is given by (\cite{skillman}) 
\begin{equation}
    Y_p = 0.234  \pm  0.002(\mbox{stat}) \pm 0.005(\mbox{syst}) 
\end{equation}

It is believed that the abundances of ${}^7$Li observed in the 
population II metal poor halo stars in our galaxy represent the
primordial
values.
 We adopt the most recent data
 (\cite{boni})\footnote{Here systematic errors due to cosmic spallation and
 depletion in stars are not included. However, even if we include such
 errors the conclusion in this paper is not changed.}:
\begin{equation}
 y_{7p} = (1.73 \pm 0.05(\mbox{stat}) \pm 0.20(\mbox{syst})) \times 10^{-10}.
\end{equation}
 Finally we discuss the constraints of D and ${}^3$He. 
The abundance of D observed in the solar neighborhood and
 inter-stellar matter give us lower limits to the primordial
 abundance, which leads to the constraints:
\begin{eqnarray}
    y_{2\odot} &\le& y_{2p}R_X, \label{y2s} \\ 
  y_{2,ism} &\le& y_{2\odot},  \label{yism}
\end{eqnarray}
where R$_{X}$ = X$_{p}$/X$_{now}$ and X is the mass fraction
 of the hydrogen ( the subscripts
 ``$ism$'', ``$\odot$''and ``$now$''
denote inter stellar matter, pre-solar, 
the present universe, respectively ).
 We also adopt a constraint
on D and ${}^3$He by using  the chemical evolution 
 model which gives a relation between the primordial and pre-solar
 abundances of D and ${}^3$He (\cite{tosi}):
\begin{equation}
 \left[ - y_{2p} + \left(\frac1{g_3} - 1 \right)y_{3p}\right]y_{2_\odot}
 - \frac1{g_3}y_{2p}y_{3_\odot} + (y_{2p}^2 + y_{3p}y_{2p})R_X
 \le 0 ,\label{stc}
\end{equation}
where $g_3$ is the survival factor of ${}^3$He, and is estimated to 
be 0.25 - 0.50 (\cite{dearborn})\footnote{The survival
 factor $g_3$ may be greater than previously thought due to the
 uncertainty in the estimation of the galactic chemical evolution 
 (\cite{schramm}), however we take the conservative values in this paper.}. The observational  abundances with
1$\sigma$ error are given by (\cite{geiss},  \cite{linsky},  \cite{tosi})
\begin{eqnarray}
  y_{2\odot} &=& (2.57 \pm 0.92) \times 10^{-5},\\
  y_{3\odot} &=& (1.52 \pm 0.34) \times 10^{-5},\\
       R_{X} &=& 1.1 \pm 0.05,\\
  y_{2,ism}  &=& (1.6 \pm 0.2) \times 10^{-5}.
\end{eqnarray}

Here we make a comment on the recent measurements in Ly-$\alpha$
absorption lines of QSOs. 
It was expected that they would provide D abundances close to the primordial
value. However two groups have reported inconsistent results on the
measurement of D/H. ($y_{2,Ly-\alpha} \sim (1.9 - 2.5) \times 10 ^{-4}$
 (\cite{songaila},  \cite{ahogan},  \cite{bhogan}). and 
$y_{2,Ly-\alpha} \sim 2.5 \times 10 ^{-4}$
 (\cite{tytlera}, \cite{tytlerb}).
 The lower value is the same order as that obtained by
the measurements in the solar system. The higher value leads to
 better agreement between theory and observation
 (\cite{olive}). At present it is premature to judge which
 result is correct. 
 Therefore we adopt neither of these values in this paper.


The lepton asymmetry $L= \sum L_l$ is defined in analogy with the baryon
asymmetry as
\begin{equation}
       L_l \equiv \frac{n_{l}-n_{\bar{l}}}{n_{\gamma}} ,
\end{equation}
where $n_{\gamma}$ is photon number, $n_l$ is number density of
lepton {\it l} ($ l = e,\mu,\tau ,\nu_e,\nu_{\mu},\nu_{\tau}$) and
$\bar{l}$ denotes the anti-particle of {\it l}. If the
lepton $l$ is in thermal equilibrium with temperature $T_l$ much
higher  than its mass, $n_l$ and $n_{\bar{l}}$ are given by
\begin{eqnarray}
       n_{l} &=& g_{l}\frac{T_{l}^3}{2\pi^2}
                       \int\frac{x^2}{1+\exp(x-\xi_{l})}dx ,\\
       n_{\bar{l}} &=&g_{l}\frac{T_{l}^3}{2\pi^2}
                       \int\frac{x^2}{1+\exp(x+\xi_{l})}dx ,
\end{eqnarray}
where $\xi_{l} = \mu_l/T_l$ denotes the
degeneracy parameter of
each lepton, $\mu_l$ is the chemical potential and  
$ g_l$ is 1 for neutrinos and 2 for
charged leptons. 
Then the energy density of each type of lepton is given by
\begin{equation}
    \rho_{l} = g_{l}\frac{T_{l}^4}{2\pi^2}
                     \int\left(\frac{x^3}{1+\exp(x-\xi_{l})} + 
                          \frac{x^3}{1+\exp(x+\xi_{l})}\right)dx.
\end{equation}
Since we only need $|\xi_{l}| \ll 1$ to solve the
discrepancy in BBN, the energy densities are expanded in $\xi_{l}$
as
\begin{equation}
    \rho_{l} = g_{l}\frac78\left(\frac{2\pi^2}{30}\right)
                     T_{l}^4\left[1+
                   \frac{30}{7}\left(\frac{\xi_{l}}{\pi}\right)^2
                     + \frac{15}{7}\left(\frac{\xi_{l}}{\pi}
                    \right)^4+\cdots\right].
\end{equation}
Since $\rho_l$ is even function of $ \xi_{l} $ and monotonically
increases, a non-zero  $ \xi_{l} $ speeds up the cosmic
expansion compared with the non-degenerate case, which leads to the 
earlier $n/p$ freeze out and more ${}^4$He production.

The lepton to photon ratio is also expanded in $\xi_{l}$ as
\begin{equation}
    L = \sum_{l}g_{l}\frac{\pi^3}{12\zeta(3)}
                  \left(\frac{T_{l}}{T_\gamma}\right)^3
                  \left[\left(\frac{\xi_{l}}{\pi}
                    \right)+\left(\frac{\xi_{l}}{\pi}
                    \right)^3+\cdots\right] 
\end{equation}
The pre-factor of the above equation is $O(1)$. Thus the lepton to
photon ratio is roughly equal to the sum of the degeneracy parameters 
$\xi_{l}$. At the epoch of the nucleosynthesis
 (T
\lower-1.2pt\vbox{\hbox{\rlap{$<$}\lower5pt\vbox{\hbox{$\sim$}}}}
 1 MeV), muons
and tau leptons have already disappeared through annihilation and decay
processes. Thus the degeneracy parameters $\xi_{\mu}$ and
$\xi_{\tau}$ are effectively zero.
 Since the neutrality of the  universe tells us 
$ (n_{e}-n_{\bar{e}})/n_{\gamma} \simeq n_B /n_{\gamma}
= \eta \sim 10^{-10} $ , $ \xi_e $ is also
negligible. Therefore  we need pay attention to only the lepton
number asymmetry of
the neutrinos. Among degeneracy parameters of three
species of neutrinos ( $\nu_e$ , $\nu_{\mu}$ , $\nu_{\tau}$ ),  
$\xi_{\nu_e}$ is the most important since the electron neutrino degeneracy
modifies the neutron to proton ratio at freeze-out. Using the
condition for chemical equilibrium, the  neutron-proton ratio $(n/p)_f$
at freeze-out is given by
$\left(\frac{n}{p}\right)_{f} = \exp( - Q/T_f - \xi_{\nu_e})$, 
where $T_f$ denotes the freeze-out temperature and $Q = 1.29$MeV. 
 Thus if $\xi_{\nu_e}$ is positive,
neutron fraction is suppressed and hence the production of ${}^4$He is 
reduced. 
 This effect is more significant than
the speed-up effect due to increase of $\rho_{\nu_e}$. 
Thus we expect that the abundance of ${}^4$He
decreases if $ \xi_{\nu_e} > 0$.
On the other hand the degeneracy
parameters $\xi_{\nu_{\mu}}$ and $\xi_{\nu_{\tau}}$ have only the speed-up 
effect which increases the $n/p$ ratio. 
Therefore large mu- or tau-neutrinos 
 degeneracies ($\xi_{\nu_{\mu}}$ or $\xi_{\nu_{\tau}}$
 $\gg$ $\xi_{\nu_e}$) might 
compensate the decrease of the $n/p$ due to electron neutrino
 degeneracy. However it seems unnatural for
 $\xi_{\nu_{\mu}}$  or $\xi_{\nu_{\tau}}$ to be much larger than
 $\xi_{\nu_e}$ , and it is easy to reinterpret the results for
 $ \xi_{\nu_{\mu}} = \xi_{\nu_{\tau}} = 0 $ in the case of large   
$\xi_{\nu_{\mu}}$  or $\xi_{\nu_{\tau}}$.
 In this paper we only study the big bang
nucleosynthesis 
with the electron 
neutrino degeneracy.

 
We use  Monte Carlo simulation and the likelihood
method to obtain the best fit values for $\xi_{\nu_e}$ and $\eta$ and
estimate their errors. We first compute the light element 
abundances and their uncertainties, using the Monte Carlo method. 
 The uncertainties come from
statistical errors in measurements of the neutron lifetime
 (\cite{group}) and nuclear
reaction rates (\cite{kawano}). We assume
 that the distribution of reaction rates is gaussian.
  Then the distributions of the calculated abundances of light
 elements also become
approximately gaussian. The theoretical 
likelihood function of the light element $ i $
is given  by
\begin{equation}
L^{th}_{i}(y_i, \xi_{\nu_e}, \eta)=\frac{1}{\sqrt{2\pi}\sigma_i^{th}
                             (\xi_{\nu_e},\eta)}\exp\left(-
                 \frac{(y_i-\mu^{th}_i(\xi_{\nu_e},\eta))^2}
                  {2\left({\sigma_i^{th}
                  (\xi_{\nu_e},\eta)}\right)^2}\right) ,
\end{equation}
where the mean value $ \mu_i $ and the variance 
$ \sigma^2_i $  depend on the
model parameters, $\eta$ and $ \xi_{\nu_e} $, and  i = 2, 3, 4, 7 
denote the D, ${}^3$He, ${}^4$He, and ${}^7$Li, respectively.

Let us estimate the observational likelihood functions. For
observational data of ${}^4$He and $ {}^7$Li we have the statistical
errors $ \sigma_{stat} $ and the systematic errors $ \sigma_{syst} $ . Here
we treat the statistical uncertainty as a gaussian distribution, and the
systematic one as a top hat or flat distribution. Namely for 
$ y_i^{obs} = \mu_i^{obs} \pm \sigma_{stat}^{obs} \pm
{}^{\sigma_{sys+}^{obs}}_{\sigma_{sys-}^{obs}} $ 
, the likelihood functions are given by
\begin{eqnarray}
\L^{obs}(y_i)
                      &=&\frac{1}{2\left(\sigma^{obs}_{sys+}
               +\sigma^{obs}_{sys-}\right)}
               \left[{\rm erf}\left(\frac{y_i-\mu_i^{obs}+\sigma_{sys-}^{obs}}
               {\sqrt{2}\sigma_{stat}^{obs}}\right)
               -{\rm erf}\left(\frac{y_i-\mu_i^{obs}-\sigma_{sys+}^{obs}}
               {\sqrt{2}\sigma_{stat}^{obs}}\right)\right],      
\end{eqnarray}
where {\rm erf(x)} is the error function.
The full likelihood function of ${}^4$He and ${}^7$Li are obtained by
convolving the theoretical and observational likelihood functions:
\begin{eqnarray}
  L_i(\xi_{\nu_e},\eta) = \int dy_iL^{th}_i(y_i, \xi_{\nu_e}, \eta)
 \times L_i^{ob}(y_i).
\end{eqnarray}  
Furthermore we regard the likelihood function as a probability
distribution function. 

For D and ${}^3$He , we take a more complicated
procedure. The statistical error is treated
as a gaussian distribution.
 Since we use the observed
 pre-solar abundances of D and ${}^3$He, we should perform the
integration over regions V in which the 
solar and galactic evolution constraints, Eqs.(\ref{y2s}),
(\ref{yism}) and 
 (\ref{stc}) are satisfied for a fixed $
 {}^3$He survival fraction. We take $ g_3 = 0.25 $ here  (\cite{crisis},
 \cite{holt}). 
Thus the likelihood function for D and ${}^3$He is given by
\begin{eqnarray}
\L_{23}(\xi_{\nu_e}, \eta)
= \int_Vdy_{2ism}dy_{2\odot}dy_{3\odot}dR_X 
       dy^{th}_2dy^{th}_3
       L_{2ism}(y_{2ism})\times L_{2\odot}(y_{2\odot}) 
        \times L_{3\odot}(y_{3\odot})\nonumber \\ \times L_{R_X}(R_X)
        \times L^{th}_2(y^{th}_2, \mu_2^{th}(\xi_{\nu_e}, \eta),
       \sigma_2^{th}(\xi_{\nu_e}, \eta) )
       \times L^{th}_3(y^{th}_3, \mu_3^{th}(\xi_{\nu_e}, \eta),
       \sigma_3^{th}(\xi_{\nu_e}, \eta) ).
\end{eqnarray}
Then the  total likelihood function is given by
\begin{equation}
       L_{tot}(\xi_{\nu_e}, \eta) = L_4(\xi_{\nu_e}, \eta)
                                \times L_{23}(\xi_{\nu_e}, \eta)
                                \times L_7(\xi_{\nu_e}, \eta).
\end{equation}
The best fit region is obtained by the $ \chi^{2} $ method.

The three likelihood functions $L_4, L_{23}$ and $L_7$  are shown 
in Fig.1
for SBBN, i.e. $\xi_{\nu_e}$ = 0. It 
is seen that these functions do not agree with each other. In
particular there is a significant difference between $L_4$ and
$L_{23}$,  which leads to the BBN crisis pointed out by 
  \cite{crisis}. 
The positive electron neutrino degeneracy reduces abundance of 
the ${}^4$He as explained before. 
 In fact in Fig.1
 (A) it is seen that the 
 Yp likelihood function moves to the
right in $\eta$ for $\xi_{\nu_e}$ = 0.05.
 On the other hand the likelihood functions
$L_{23}$ and $L_7$ are not changed very much since 
they are almost independent of the
neutron abundance at freeze-out. Then the four likelihood functions agree
with each other and the discrepancy in SBBN is solved (Fig. 2).
In Fig.3
 we show the contours of the confidence levels in the 
$\eta$-$\xi_{\nu_e}$  plane. The best fit values for $\eta$ and
$\xi_{\nu_e}$ are estimated as 
\begin{eqnarray}
 \eta_{10} &=& \;4.0 \;^{+1.5}_{-0.9}  \qquad\qquad\quad
                   \left(95 \% \;\mbox{C.L.}\right) ,\\
 \xi_{\nu_e} &=& \left(4.3 \;^{+4.0}_{-4.0} \right)\times 10 ^{-2} 
           \quad \left(95 \% \;\mbox{C.L.}\right) ,
\end{eqnarray}
where $\eta_{10} = \eta \times 10^{10}$. Notice that standard BBN
( $\xi_{\nu_e}$ = 0) is outside of the 95$\%$ C.L. contour.
Though we adopt a lower
$g_3$ here, in the case of a higher survival factor ($g_3$ = 0.5) the
result is not changed very much, see Fig.2.

\section{Conclusion}
In this paper we have investigated BBN with lepton number
asymmetry. Assuming that the electron neutrino has a significant chemical
potential due to lepton asymmetry, it has been shown that the 
observational data  agree with the theoretical predictions and the BBN
crisis is solved. We have 
estimated the lepton number density of electron neutrinos
  and the baryon number
 density. The  baryon density
 parameter is given by 
$\Omega_bh^2 = 0.015\;^{+0.006}_{-0.003}$
 (95 $\%\;$ C,L.), where $h$ is $H_{0}$ = 100 $h$
 km/s/Mpc. The estimated
 chemical potential of $\nu_e$ is about $10^{-5}$ eV which is much
 smaller than experiments can detect ($\sim$ 1 eV, \cite{kon}). In other
 words, BBN gives the most stringent constraint on the chemical
 potential of $\nu_e$.

\acknowledgments
We thank P. Kernan for helpful advice about BBN code, and T.Totani for 
valuable discussions about statistical method. This work has been
supported in part by the Grant-in-Aid for COE Research (07CE2002) of
the Ministry of Education, Science, and Culture in Japan  
, and in part by Grant-in-Aid for Science Research Fund of the Ministry of
Education, Science, and Culture No.08740194.

\newpage
\figcaption[three]{(A) Likelihood functions of Y$_P$ (${}^4$He mass
    fraction), (B) likelihood functions of D \& ${}^3$He for $g_3 =$
    0.25, 0.50, 
      (C) likelihood functions of ${}^7$Li.
      The solid line represents for $\xi_{\nu_e}$ = 0.00 (SBBN) and the
      dashed line represents for  $\xi_{\nu_e}$ = 0.05. For ${}^4$He
       with an electron
      neutrino degeneracy, the peak of the function
      moves to the right in $\eta$ axis.
      The intensity is an arbitrary
    scale. \label{three}}

\figcaption[yield]{The contours of the confidence levels in
             ( abundance $-$ $\eta$ ) plane for each light elements 
             ${}^4$He, D and ${}^7$Li for $\xi_{\nu_e} = 0.05$. 
            In the contours the solid line denotes
             68$\%$C.L. and the dotted line denotes 95$\%$C.L..
             Theoretical predictions are also shown with theoretical
             uncertainties (1 $\sigma$)
             obtained by Monte Carlo method (for $g_3$ =
             0.25). \label{yields} }

\figcaption[cont]{The contours of the confidence
     levels in the parameter space,
    $ \eta $  -  $ \xi_{\nu_e} $ for $g_3$ = 0.25, 0.50. The dashed
    (solid) contour
    corresponds to 68$\%$ C.L. (95$\%$ C.L.).
      Standard BBN ( $\xi_{\nu_e}$ = 0.00) is outside of
      the 95$\%$ C.L.. \label{cont}}




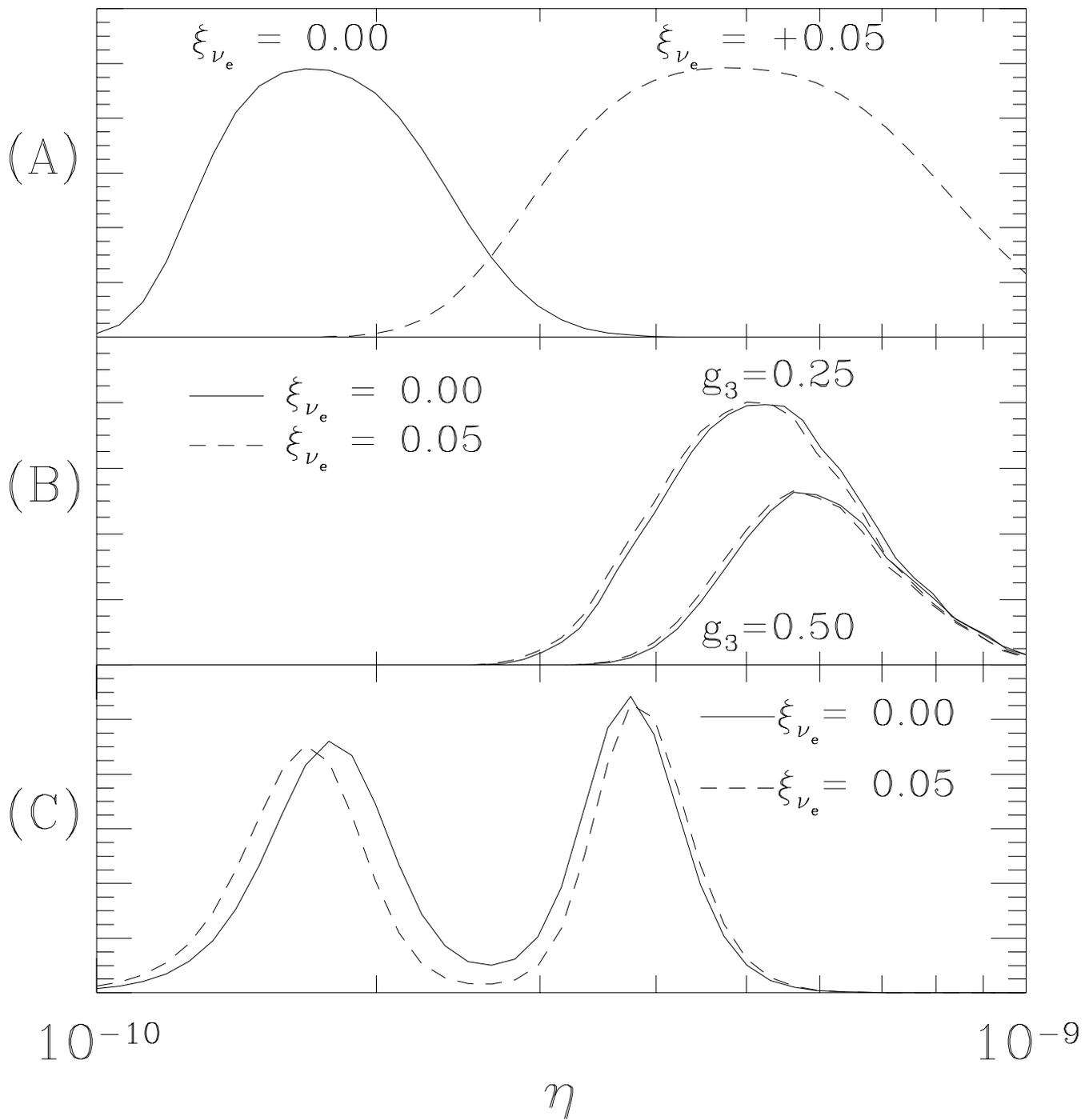
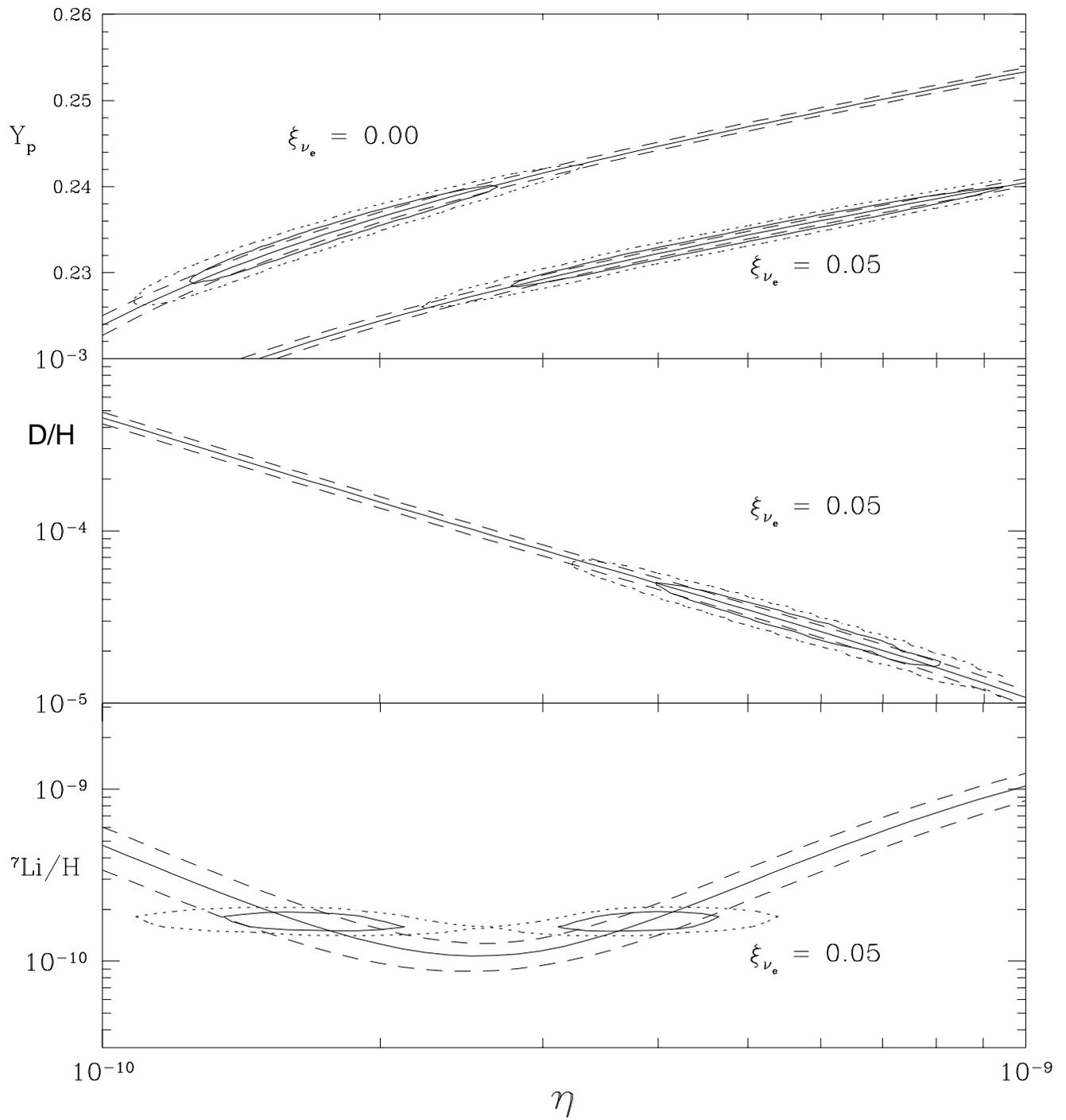
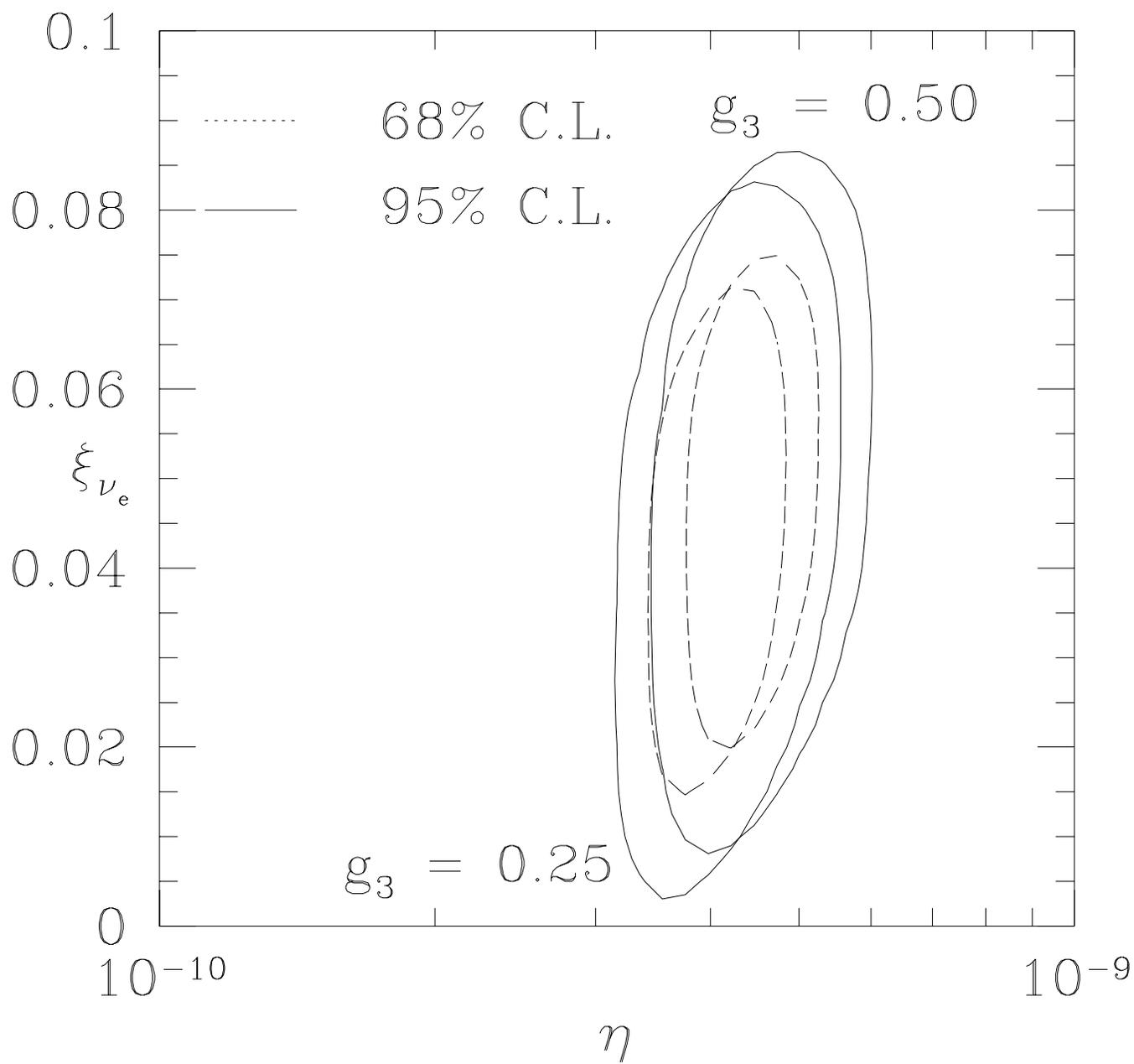
\end{document}